\documentclass[prl,floatfix,twocolumn,showpacs]{revtex4}
\usepackage{amsmath}
\usepackage{amsfonts}
\usepackage{mathrsfs}
\usepackage{epsfig}

\usepackage{graphicx}
\usepackage{dcolumn}
\usepackage{amsmath}

\begin{document}
\title{
Wave-train Induced Unpinning of Weakly Anchored Vortices in Excitable Media
}
\author{Alain Pumir$^{1,2}$, Sitabhra Sinha$^3$, S. Sridhar$^3$,
M\'ed\'eric Argentina$^2$, Marcel H\"orning$^4$, Simonetta Filippi$^5$, 
Christian Cherubini$^5$, 
Stefan Luther$^6$ and Valentin Krinsky$^{6,7}$ }
\affiliation{%
$^1$Laboratoire de Physique, ENS de Lyon and CNRS, 46 All\'ee d'Italie, 69007, Lyon, France.}
\affiliation{%
$^2$Lab J. A. Dieudonn\'e, Universit\'e de Nice and CNRS, Parc Valrose, F-06108 Nice Cedex, France.}
\affiliation{%
$^3$The Institute of Mathematical Sciences, CIT Campus, Taramani,
 Chennai 600113, India.}
\affiliation{%
$^4$ Dept. of Physics, Graduate School of Science, Kyoto University,
Kyoto 606-8502, Japan.}
\affiliation{%
$^5$Nonlinear Physics \& Math Modeling Lab, University Campus Bio-Medico, I-00128, Rome, Italy.}
\affiliation{%
$^6$MPI for Dynamics and Self-Organization,  Am Fassberg 17, D-37077, G\"ottingen, Germany.}
\affiliation{%
$^7$Institut Non Lin\'eaire de Nice and CNRS, F-06560, Valbonne, France.
}%
\date{\today}
\begin{abstract}
A free vortex in excitable media can be displaced and removed by
a wave-train. However, simple physical
arguments suggest that vortices anchored to large
inexcitable obstacles cannot be removed similarly. We show that unpinning
of vortices attached to obstacles smaller than the core radius of the
free vortex is possible through pacing. The wave-train frequency necessary for 
unpinning increases with the obstacle size and we present a
geometric explanation of this dependence. Our model-independent results suggest that
decreasing excitability of the medium can facilitate pacing-induced
removal of vortices in cardiac tissue.
\end{abstract}

\maketitle

\newpage
Rotating spiral waves of propagating excitation characterize the disruption
of ordered behavior in excitable media describing a broad class of
physical, chemical and biological systems~\cite{CrossHoh93}. 
In the heart, spiral waves of electrical activity have
been associated with life-threatening arrhythmias~\cite{VIK80,Davidenko92,Zipes04},
i.e., breakdown of the normal rhythmic pumping action of the heart.
Controlling such spatial patterns
with low-amplitude external perturbation is a problem of fundamental interest
\cite{Sinha01,T04,ZCWYH05,Sinha08,Isomura08} with significant implications 
for the clinical treatment of cardiac arrhythmias~\cite{Fenton09}.

In a homogeneous active medium, a spiral wave can be controlled by a 
wave train, induced by periodic stimulation from a local source
(pacing)\cite{Zipes04}. If the
frequency of stimulation is higher than that of the spiral wave, the wave
train induces the spiral to drift. 
In a finite medium, the vortex is
eventually driven to the boundary and thereby
eliminated from the system~\cite{Agladze83,Agladze07,GPK01}.
Inhomogeneities in the medium, such as inexcitable obstacles, can anchor the
spiral wave preventing its removal by a stimulated wave-train \cite{Pertsov84}.
This mechanism is analogous to pinning of vortices in disordered
superconductors~\cite{Blatter94,Pazo04}.
In the heart, 
obstacles such as blood vessels or scar tissue, 
can play the 
role of pinning centres~\cite{Tung06}, leading to anatomical reentry, 
the sustained periodic excitation of the region around the obstacle. 

In the immediate neighborhood of the obstacle, pinned vortices 
are qualitatively equivalent to waves circulating in a one-dimensional ring.
They can be removed by external stimulation provided the electrode is located
on the reentrant circuit, i.e., the closed path of the vortex around
the obstacle, and the stimulus is delivered within a 
narrow time interval~\cite{Glass95}.
However, for the more general situation of
pacing waves generated far away from the reentrant circuit,
a classical result due to Wiener and Rosenblueth (WR) states that,
all waves circulating around such obstacles are created or annihilated in 
pairs (see Ref.~\cite{Wiener}, in particular, pp.216-224). This implies
that it is impossible to unpin the spiral wave by a stimulated wave train. 

In this paper, we demonstrate that the
WR mechanism for the failure of pacing in unpinning spiral waves is
valid only when the radius of the {\em free} spiral core 
(i.e., the closed trajectory of the spiral tip defined as
a phase singularity~\cite{Winfree87})
is small compared to the size of the obstacle. 
We elucidate the transition between the case
of a free vortex and one attached to a large obstacle by systematically
decreasing the core radius of the free spiral, $R_{FS}$, relative to the 
obstacle size, $R_{obst}$, by
increasing the excitability of the medium. 
Our main result is that an anchored rotating wave can be removed by 
a stimulated wave train provided $R_{FS} > R_{obst}$.

To illustrate our arguments, we use the simple model of excitable media 
introduced in Ref.~\cite{Bark90},
described by an excitatory ($u$) and a recovery ($v$) variable:
\begin{eqnarray}
\partial_t u & = & \frac{1}{\epsilon} u ( 1 - u) [ u - ( v + \frac{b}{a} ) ] 
+ \nabla^2 u, \nonumber\\
\partial_t v & = & ( u - v ), \label{Barkley}
\end{eqnarray}
where, $a$ and $b$ are parameters describing the kinetics. The relative
time-scale $\epsilon$ between the local dynamics of $u$ and $v$ is set
to $0.02$.
We discretize the system on a square spatial grid of size $L \times
L$, with a lattice spacing of
$\Delta x = 0.25$ and time step of $\Delta t = 0.01$. For our simulations 
$L= 200$.
We solve Eq.~\ref{Barkley} using forward Euler scheme with a standard
nine-point stencil for the Laplacian.
No flux boundary conditions are implemented 
at the edges of the simulation domain. An obstacle is implemented by introducing
a circular region of radius $R_{obst}$ in the center of simulation
domain inside which diffusion is absent.
Pacing is delivered by setting the value of $u$ to $u_p = 0.9$ 
in a region of $6 \times 3$ points at the center of the upper boundary of 
the simulation
domain. The maximum pacing frequency is limited by the {\em refractory period},
$T_{ref}$, the duration for
which stimulation of an excited region does not induce a response. 

\begin{figure}
\centering
\includegraphics[width=0.98\linewidth,clip]{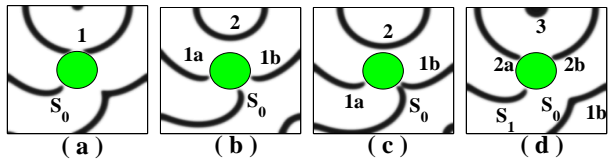}
\caption{
(a) Wave $S_0$, pinned to an obstacle (shaded), 
rotates counterclockwise; wave $1$ is the first pacing wave.
(b) Wave $1$ hits the obstacle, and 
separates into a wave rotating counterclockwise ($1a$) 
and a wave rotating clockwise ($1b$).
(c) Waves $S_0$ and $1b$ collide and merge
leaving only one rotating wave $1a$ denoted $S_1$ hereafter.
(d) The wave resulting from the merging of $S_0$ and $1b$ leaves the 
system. 
The interaction between the following pacing wave, $2$ and $S_1$, is 
similar to that shown in (a-c). 
Thus, the pinned vortex persists. 
Numerical simulation of the Barkley model with parameters:
$a = 0.9$, $b=0.17$;
the pacing 
period is $T_p = 6.7$ and the radius of the obstacle is $R = 6.5$.
}
\label{Parity_Wiener}
\end{figure}
When the obstacle size is large relative to the core radius
of the free spiral, $R_{FS}$, 
the failure of a wave train in unpinning the vortex is
illustrated in Fig.~\ref{Parity_Wiener}.
Initially, the spiral wave $S_0$ rotates counterclockwise around the obstacle.
During the interaction with pacing waves, the number of waves attached to
the obstacle
can change due to two possible processes (see Ref.~\cite{Wiener}, p.216 
and 220).
First, when the pacing wave reaches the obstacle, 
it splits into two oppositely rotating waves:
one clockwise and the other counterclockwise.
Second, collision between two rotating 
waves, as seen in Fig.~\ref{Parity_Wiener}(c), results in the annihilation
of a pair of counterclockwise and
clockwise waves.
In both cases, the number of waves rotating counterclockwise is always 
larger than the number rotating clockwise by $1$. 
Thus, in addition to conservation of total topological charge (i.e., sum of 
the individual chiralities, $+1$ or $-1$) for {\em all} spiral waves in a 
medium~\cite{Glass77,Winfree87}, 
topological charge {\em around} the obstacle also appears to be conserved. 
However, in the limiting case of infinitesimally small obstacle
corresponding to a free vortex, a stimulated wave-train with frequency
higher than that of the spiral wave
will always succeed in displacing the latter, eventually
removing it from a finite medium. Thus, there is a transition from failure to
successful pacing as $R_{obst}$ is reduced relative to $R_{FS}$.

\begin{figure}
\centering
\includegraphics[width=0.98\linewidth,clip]{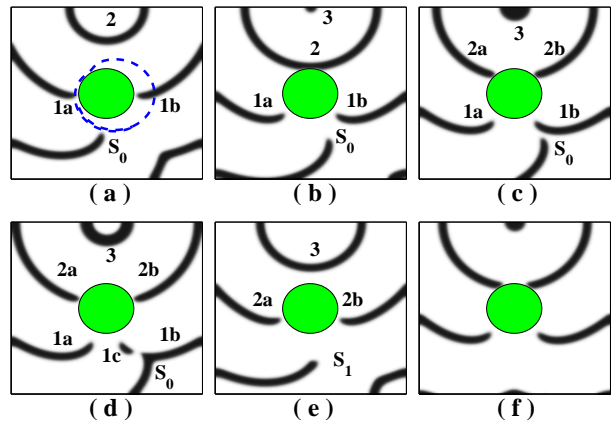}
\caption{
Lowering excitability results in successful detachment of
pinned vortex by pacing.   
$S_0$ is a rotating wave whose core (dashed line) is larger than
the pinning center (shaded). 
(a-c) are topologically as in Fig.~1.
(d)  A wavelet  $1c$ is produced after collision of waves  $S_0$   
and $1b$, in contrast with Fig.~1(d).
(e) The wavelet $1c$ collides with $1a$ and 
the resulting wave $S_1$ is displaced away from the obstacle.
(f) Subsequent pacing induces drift of the spiral wave $S_1$ to the
boundary, eventually removing it from the medium.
The parameters are as in Fig.~\ref{Parity_Wiener}, except for $a=0.895$
and $b=0.1725$, resulting in
increasing the vortex core size. 
}
\label{Spiral_Detach}
\end{figure}
The primary fact responsible for this
transition is that the spiral wave is no longer in physical 
contact with an obstacle of size smaller than
$R_{FS}$ \cite{Tung06}, contrary to the fundamental assumption of 
Ref.~\cite{Wiener}.
Fig.~\ref{Spiral_Detach} shows an explicit example
of successful detachment of a pinned wave from the obstacle boundary,
where the core radius of 
a free spiral in the medium is 
made larger than $R_{obst}$ by diminishing the excitability
of the system.

The possibility of unpinning the wave in Fig.~\ref{Spiral_Detach}
can be traced to the following fact:
the collision between $S_0$ and the pacing wave-branch $1b$ occurs 
{\em a small distance away} from the obstacle boundary and does not result
in complete annihilation of both waves.
A small fragment $1c$ survives in the spatial interval between 
the collision point and the
obstacle [Fig.~\ref{Spiral_Detach}(d)].
If the tip of $S_0$ is close to the obstacle, the fragment $1c$
is small, and rapidly shrinks and disappears.
However, if the gap between the reentrant wave tip and the obstacle is large
at the collision point, such that the size of $1c$ is larger than
a critical value $l_n$, the fragment can survive.
As $1c$ propagates further away, it
collides with the pacing wave $1a$ and forms a new broken wave $S_1$ that is
completely  detached from the obstacle.
Interaction with successive pacing
waves progressively pushes the vortex further away from the obstacle, and 
eventually from a finite medium.
The difference between the number of 
spirals rotating counterclockwise and 
clockwise {\it around the obstacle} changes from $1$ initially
(Fig.~\ref{Spiral_Detach}, a), to $0$ in Fig.~\ref{Spiral_Detach}(e), 
contrary to what happens for a larger obstacle (Fig. ~\ref{Parity_Wiener}). 
The absence of topological charge conservation for waves rotating {\em around} 
a smaller obstacle underlines the breakdown of the fundamental assumption
behind the WR argument for why pacing cannot detach pinned waves.  
The unpinned wave is subsequently driven outside
the system boundaries by pacing (Fig.~\ref{Spiral_Detach}, f), 
thus eventually also reducing the total 
topological charge of the {\em finite} medium to 0.

\begin{figure}
\includegraphics[width=0.6875\linewidth,clip]{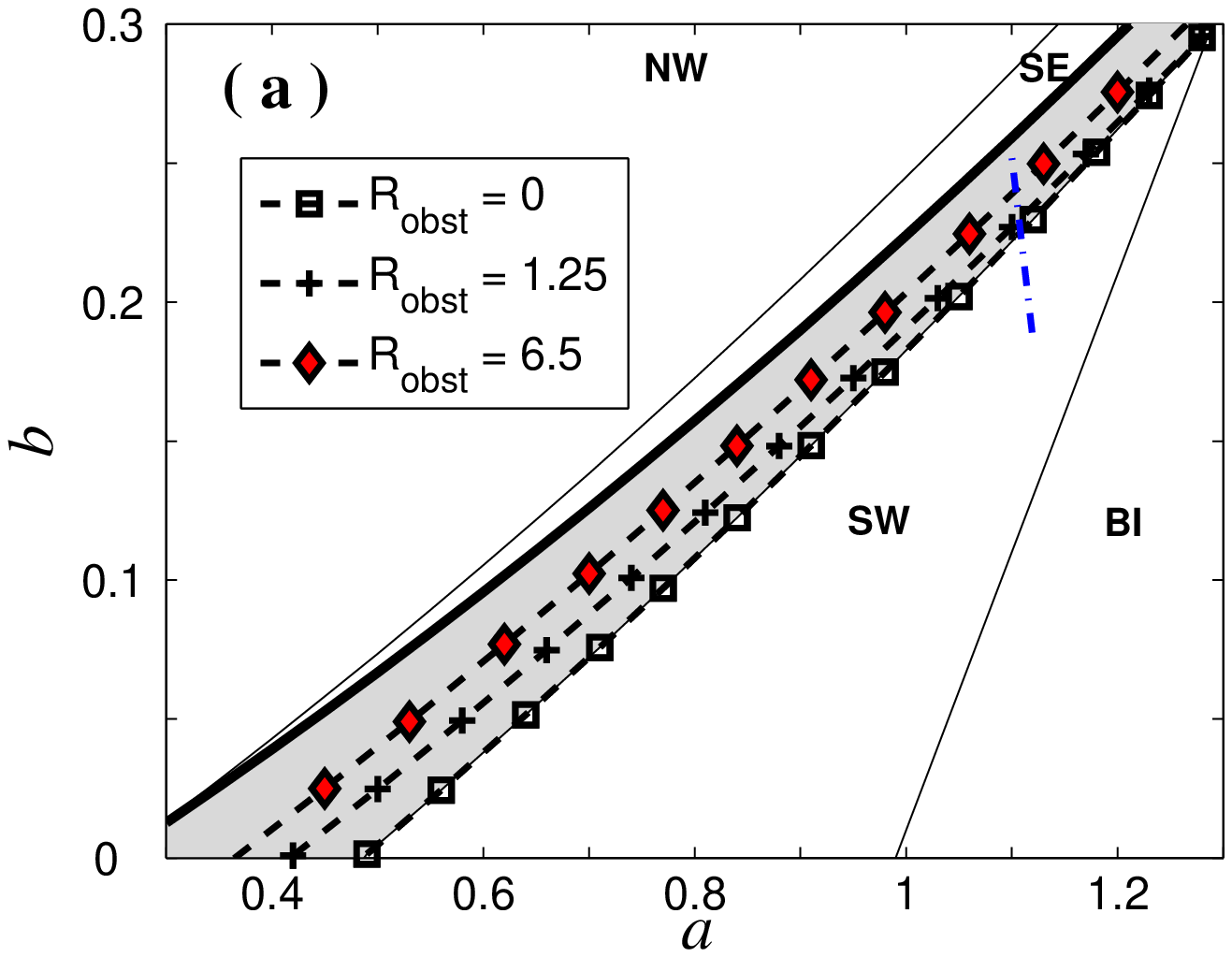}
\includegraphics[width=0.2925\linewidth,clip]{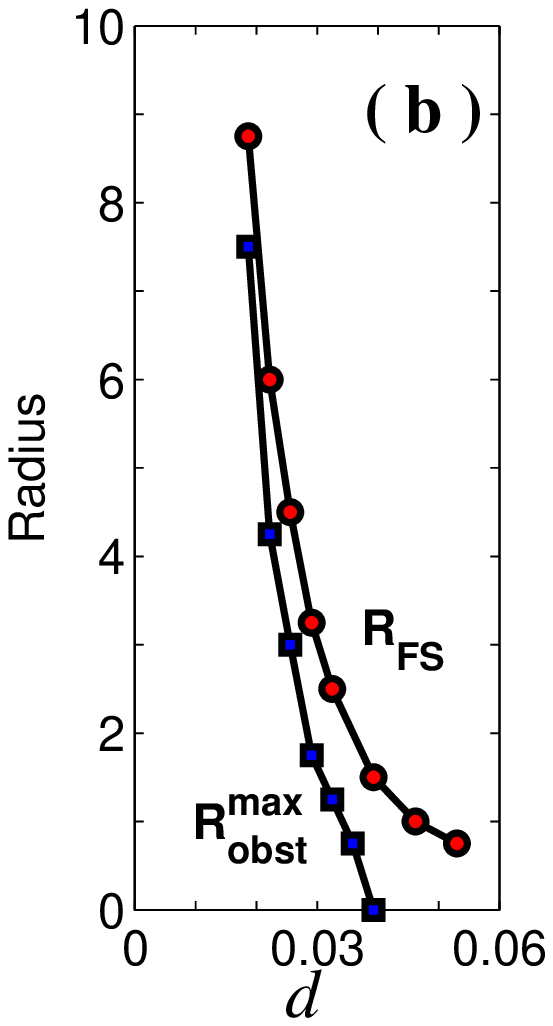}
\centering
\caption{
(a) Parameter space of the Barkley model. 
Unpinning is possible in the shaded portion of the $SW$ region, which 
exhibits persistent spiral waves. The thick line indicates the boundary
with the SE region, where spirals cannot form. The 
domain where unpinning is possible shrinks with increasing size of the 
pinning center, 
the three dashed lines corresponding to $R_{obst}=0$, i.e., no 
obstacle (square), 
$R_{obst} = 1.25$ (plus) and $R_{obst}=6.5$ (diamond).
(b) Radius $R_{FS}$ of the free spiral  
and the maximum obstacle radius  $R_{obst}^{max}$ 
from which 
wave trains can unpin vortices,
as a function of the 
distance $d$ from the SE-SW boundary, 
along the dot-dashed line indicated in (a). 
Note that $R_{FS} > R_{obst}^{max}$, and both increase with decreasing $d$.
[In (a), NW (BI) indicates the parameters for which steady 
waves are absent (the medium is bistable).]
}
\label{Bark_param}
\end{figure}
The relative size of the obstacle, compared to the free spiral core,
is the key parameter that decides whether a pinned reentrant wave can 
be removed or not. Indeed,
the radius of the free spiral core in the successful case,
$R_{FS} = 9.05$ (Fig. ~\ref{Spiral_Detach}) is
significantly larger than in the unsuccessful one, 
$R_{FS} = 5.80$ (Fig. ~\ref{Parity_Wiener}). It is further confirmed
by a detailed numerical study of the interaction between a pacing wave train
and a pinned spiral over the ($a,b$) parameter space of the Barkley model.
As shown in Fig.~\ref{Bark_param}(a), the rotating wave anchored to 
the obstacle
can be removed by pacing only in the neighborhood of the
sub-excitable (SE) region (using the terminology of Ref.~\cite{Alonso03}), 
where
$R_{FS}$ diverges [Fig.~\ref{Bark_param}(b)]. 
This is explained by noting that in the SE
regime, the tangential velocity of a broken wavefront is negative,
thus causing the front to shrink and not form a spiral. 
As we approach the regime where spiral waves are persistent (SW), 
the tangential velocity of the
wave break gradually increases to zero and becomes positive on crossing 
the SE-SW boundary, so that the broken wave front can now evolve into
a spiral. As $R_{FS}$ increases with decreasing tangential velocity
of the wave front, the spiral core becomes large 
close to the SE region resulting in successful pacing-induced termination
of pinned reentry.

\begin{figure}
\centering
\includegraphics[width=0.98\linewidth,clip]{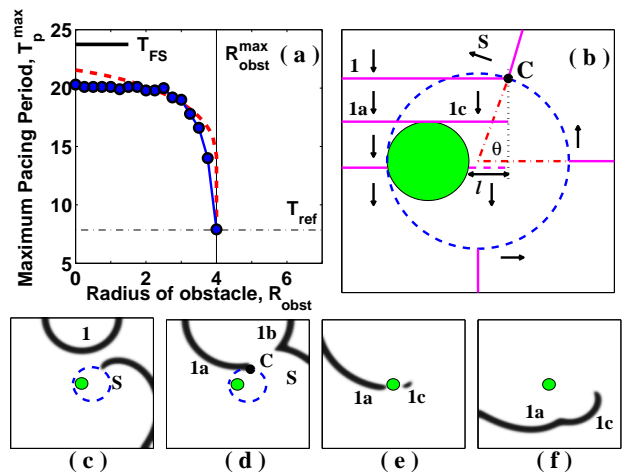}
\caption{
(a) The maximum pacing period $T_p^{max}$ at which unpinning 
is possible as a function of the obstacle radius $R_{obst}$. 
For the parameters $a = 1.1323, b = 0.2459$ that we have 
used, the maximum radius of obstacle from which depinning can occur is $R_{obst}^{max} = 4$.
$T_{FS}$  is  period of a free spiral wave and $T_{ref}$ is the refractory period. 
The 
dashed line indicates the prediction from Eq.~\ref{eq2}.
(b) The wavelet formation mechanism leading to the detachment of the
pinned vortex (schematic). (c-f) Numerical 
simulation of the Barkley model.
$S$ collides with wave $1$ at point $C$ at an angle $\theta$.
The part $1b$ of the pacing wave merges with $S$, moving out of the system. 
The remaining part of the pacing wave collides with the obstacle (shaded) 
separating into $1a$ and a small wavelet $1c$. 
When the length $l$ of wavelet $1c$ is larger than the critical nucleation 
length, $1c$ survives and collides with $S$. This
results in unpinning of $S$.
}
\label{Small_Obst}
\end{figure}

We observe that there is a maximum radius of the obstacle ($R_{obst}^{max}$) 
close to $R_{FS}$ above which
pacing is unsuccessful in detaching the anchored spiral wave
[Fig.~\ref{Bark_param}(b)]. 
Fig.~\ref{Small_Obst}(a) shows that the pacing period for 
successful unpinning from the obstacle is bounded
by the refractory period ($T_{ref}$)
and a maximum value $T_p^{max}$ that
is independent of $R_{obst}$ for small obstacles.
As we approach $R_{obst}^{max}$, the upper bound sharply decreases, 
becoming equal
to the refractory time at $R_{obst}^{max}$, which indicates that pacing will be
unsuccessful in unpinning waves attached to obstacles of radii larger than
$R_{obst}^{max}$. 
Thus, the results shown in Figs.~\ref{Bark_param}(b) and ~\ref{Small_Obst}(a)
demonstrate our earlier assertion that pacing induced removal
of anchored waves will be possible only when the obstacle is 
smaller than the core radius of the free spiral wave in the medium.

Our numerical results indicate that the maximum pacing period necessary
for detaching a pinned spiral wave is a decreasing function of the
obstacle size [Fig.~\ref{Small_Obst}(a)].
This can be explained semi-quantitatively by the following geometric argument,
valid when the size of the obstacle is small compared to the core size of the 
spiral, and supported by the simulations shown in Fig.~\ref{Small_Obst}(c-f).
The tip of the spiral $S$ moves along its circular
trajectory, shown by the broken line in Fig.~\ref{Small_Obst}(b), and
interacts with the pacing wave coming from the top, represented by a solid
line. 
The part $1b$ of the pacing wave collides with $S$ at the
point $C$ characterized by 
an angle $\theta$ that the spiral tip makes with the symmetry axis
(i.e., the line joining the centers of the obstacle and
spiral core); the resulting wave eventually leaves the system 
[Fig.~\ref{Small_Obst}(d)].
The remaining section of the pacing wave splits into two
waves, $1a$ and $1c$, propagating along either side of the obstacle.
The wave tip moves approximately in a straight line from $C$, so that the length
of the wave $1c$ at the symmetry axis is
$l = R_{FS} (1 + \cos \theta ) - 2 R_{obst}$.
When the fragment $1c$ is larger than the nucleation size $l_n$, it expands 
into a wavefront that reconnects with wave $1a$. This results in
a displacement of the wave $1a$ away from the obstacle, leading to 
unpinning (as in
Fig.~\ref{Spiral_Detach}). For $l < l_n$, $1c$ shrinks and 
eventually disappears, resulting in unsuccessful pacing. 

Thus, the condition for detachment is $l \ge l_n$. The length $l$ is a 
decreasing function of the angle $\theta$, which in turn, is a decreasing
function of the pacing period, $T_p$, as explained below.
The relation between $T_p$ and $\theta$ can be established by
estimating the time interval for two successive collisions of the spiral with
the pacing waves.
From the point of collision $C$,
the pacing wave reaches the obstacle
after time
$T_1 = (R_{FS} \sin~\theta - R_{obst}) /v$, and the symmetry axis after
time $T_2 = T_1 + (R_{obst} T_{FS}/4 R_{FS})$.
From the symmetry axis, 
the new reentrant wave $S$ moves by an angle $(\theta + \pi)$ to arrive
at $C$
at time $T_3 = T_2 + [T_{FS} (\theta + \pi) / 2\pi]$, where it collides
with the next pacing wave. Noting that $T_3 = T_p$
allows us to implicitly
express $T_p$ as a function of $\theta$, and thereby, $l$.
The maximum pacing period leading to detachment is obtained when
$l = l_n$, as:
\begin{equation}
T_p^{max} = \frac{R_{FS}}{v} (\sin \theta_c - f_R) + \frac{f_R T_{FS}}{4} + 
\frac{T_{FS}(\theta_c + \pi)}{2 \pi},
\label{eq2}
\end{equation}
where, $\theta_c = \arccos (2 f_R - 1 + [l_n/R_{FS}])$ 
and $f_R=R_{obst}/R_{FS}$.
When 
$R_{obst} > R_{obst}^{max} = R_{FS} - (l_n/2)$, $T_p^{max}$ 
has complex values, indicating that for larger obstacles the fragment is 
too small to survive. 
The nucleation length $l_n$ can thus be estimated from $R_{obst}^{max}$, 
which allows us, in turn, to determine 
the dependence of $T_p^{max}$ as a function of $R_{obst}$ from Eq.~\ref{eq2}.
Fig.~\ref{Small_Obst}(a) shows this to be in fair agreement with our
numerical simulations.

We stress that the arguments used here are model independent, 
and are based only on the property
that waves in excitable media annihilate on collision.
We verified numerically that wave-train induced unpinning is
observed also in a more detailed and realistic description of
cardiac tissue, the Luo-Rudy model 
~\cite{Luo91} [see the movie on line], under conditions of reduced 
excitability. 
Meandering, 
which occurs in the Barkley model at low $a, b$ values
(Fig.~\ref{Bark_param}, a),
does not affect the physical effect discussed here.
Note that the
proposed unpinning mechanism 
is for the case of an obstacle
smaller than the vortex core. It is possible under certain circumstances to
unpin waves from obstacles larger than the core
because of other effects such as the presence of
slow conduction regions~\cite{Sinha02} and nonlinear 
wave propagation (alternans)~\cite{BS05}.

Our results thus predict that in cardiac tissue, the removal of spiral waves 
pinned to a small obstacle by high-frequency wave trains is facilitated
by decreasing the excitability of the medium. This is consistent with 
previous experimental results on cardiac preparations using Na-channel
blockers~\cite{Tung06} and our prediction could be directly tested in a similar
experimental setup~\cite{Tung06,Agladze07}.

In conclusion, we have shown that for a pinned vortex
interacting with a pacing wave train, unpinning is possible when the
size of the obstacle is smaller than that of the spiral core.
The minimum wave train frequency necessary for unpinning in the presence 
of an inexcitable obstacle is higher than that for inducing drift in
a free vortex towards the boundaries,
and it increases with the size 
of the pinning center.
Our results suggest that lowering the excitability of the medium makes 
it easier to unpin vortices by pacing. 

This research was initiated at the Kavli Institute for Theoretical Physics, and
was supported in part by IFCPAR (Project 3404-4) and IMSc Complex Systems
Project (XI Plan).


\end{document}